\documentclass[%
aip,
amsmath,amssymb,
reprint,%
]{revtex4-2}
\usepackage{graphicx}
\usepackage{dcolumn}
\usepackage{mathrsfs}
\usepackage{bm}
\usepackage{lineno}
\usepackage{hyperref}
\usepackage{bbm}
\usepackage{ulem}
\usepackage{xcolor}

\begin{document}

\title{Quantum entangled Sagnac interferometer}

\author{Wen Zhao}
\affiliation{College of Precision Instrument and Opto-Electronics Engineering, Key Laboratory of Opto-Electronics Information Technology, Ministry of Education, Tianjin University, Tianjin 300072, P. R. China}%

\author{Xuan Tang}
\affiliation{Department of Physics, City University of Hong Kong, 83 Tat Chee Avenue, Kowloon, Hong Kong, P. R. China}

\author{Xueshi Guo}
\affiliation{College of Precision Instrument and Opto-Electronics Engineering, Key Laboratory of Opto-Electronics Information Technology, Ministry of Education, Tianjin University, Tianjin 300072, P. R. China}%

\author{Xiaoying Li}
 \email{xiaoyingli@tju.edu.cn}
\affiliation{College of Precision Instrument and Opto-Electronics Engineering, Key Laboratory of Opto-Electronics Information Technology, Ministry of Education, Tianjin University, Tianjin 300072, P. R. China}%

\author{Z. Y. Ou}
 \email{jeffou@cityu.edu.hk}
\affiliation{Department of Physics, City University of Hong Kong, 83 Tat Chee Avenue, Kowloon, Hong Kong, P. R. China}

\begin{abstract}

SU(1,1) interferometer (SUI) is a novel type of interferometer that uses directly entangled quantum fields for sensing phase change. For rotational sensing, Sagnac geometry is usually adopted. However, because SUI depends on the phase sum of the two arms, traditional Sagnac geometry, when applied to SUI, will result in null  signal. In this paper, we modify the traditional Sagnac interferometer by nesting SU(1,1) interferometers inside. We show that the rotational signal comes from two parts labeled as ``classical" and ``quantum", respectively, and the quantum part, where quantum entangled fields are used for sensing, has rotational signal enhanced by a factor related to the gain of the SUI. 

\end{abstract}

\maketitle

Optical rotational sensing has application in navigation as an optical gyroscope and is usually done with a Sagnac interferometer \cite{sag,gyro} through phase measurement. Squeezed states can be injected from the unused port for the enhancement of sensitivity \cite{caves,xiao,schn,gr20}. Recently, a novel type of quantum interferometer known as the SU(1,1) interferometer (SUI)
was realized \cite{yur,apl,nc}, which utilizes nonlinear parametric processes to replace beam splitters as wave splitting and mixing elements for interference. It has been shown that SUIs possess some unique properties that are different from linear beam splitter-based traditional interferometers \cite{ou20} because the fields generated in parametric processes are quantum entangled. Particularly, they have quantum enhancement in signal-to-noise ratio (SNR) for precision phase sensing and are tolerant to external losses, which is extremely desirable for quantum sensing. Rotational sensing relies on phase measurement in a Sagnac interferometer. Thus, it is tempting to apply the ideas in SU(1,1) interferometers to Sagnac geometry for sensitivity enhancement. 
  
\begin{figure}
\centering
\includegraphics[width=0.95\linewidth]{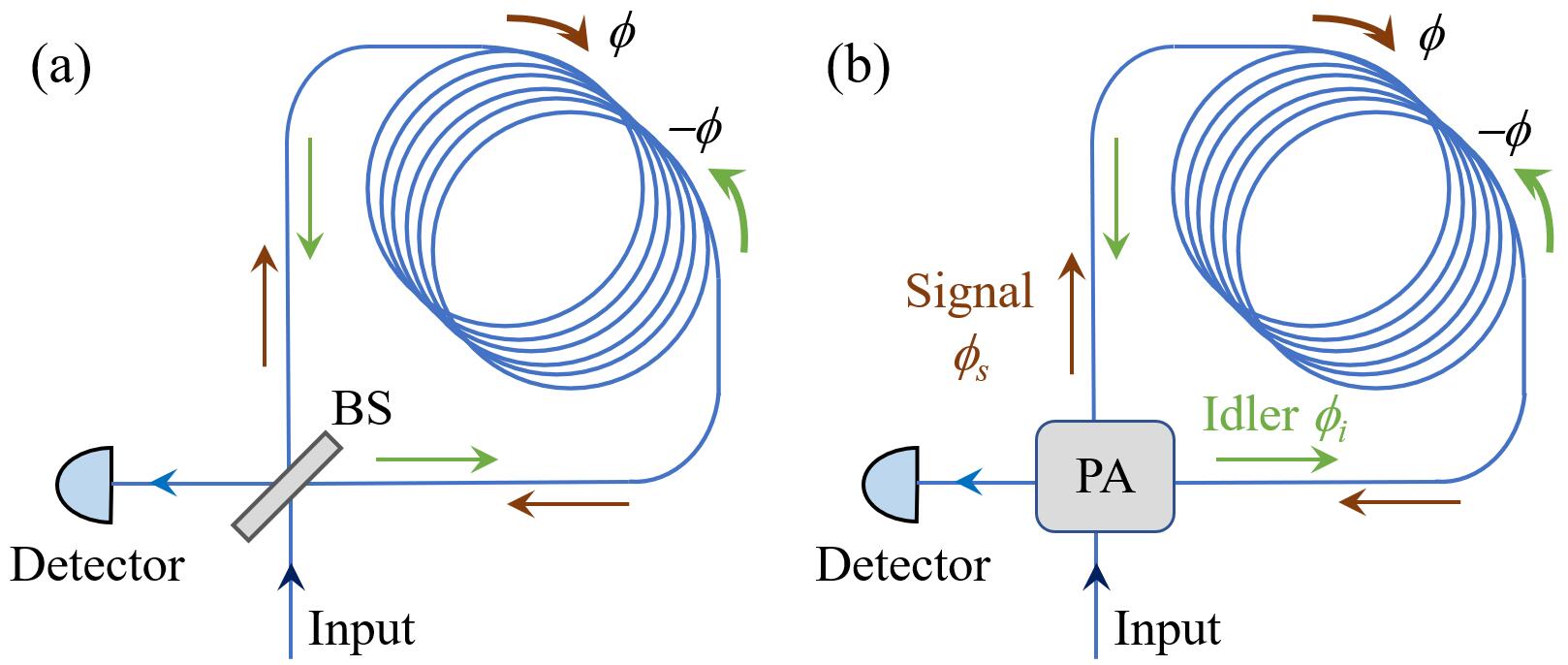}
\caption{(a) A Sagnac interferometer. (b) A simple variant of Sagnac interferometer with the beam splitter (BS) replaced by a parametric amplifier (PA) to form SU(1,1) interferometers (SUI). This scheme  is insensitive to rotation.}
\label{setup}
\end{figure}

However, a simple replacement of the beam splitter by a parametric amplifier in Sagnac interferometer will not work (Fig.\ref{setup}a and b). This is due to another unique property of SUI that  interference fringes depend on the phase sum of the two arms instead of phase difference \cite{ou20}:
\begin{eqnarray}\label{out}
I_{out}^{SUI} \propto 1+{\cal V} \cos (\phi_s+\phi_i),
\end{eqnarray}
where $\phi_{s,i}$ are the phase shifts of the signal and idler fields, respectively, and ${\cal V}$ is the visibility of interference. Since rotation in the Sagnac geometry will induce phase shifts of the same value but opposite sign in the light fields propagating in two opposite directions ($\phi$ in $\phi_s$ but $-\phi$ in $\phi_i$ in Fig.\ref{setup}(b)), the phase signal due to rotation will be canceled in the output in Eq.(\ref{out}), making it insensitive to rotation. So, it is not straightforward to implement Sagnac geometry in SU(1,1) interferomters. In this paper, we consider the phase sum property of SUI and construct a combination of SUI and Sagnac interferometer accordingly. We will show that this new configuration gives rise to rotational signal from two loops labeled as ``classical" and ``quantum". The classical loop is the same as before but the quantum loop has the rotation sensitivity enhanced by quantum entanglement.

\begin{figure}
\centering
\includegraphics[width=0.9\linewidth]{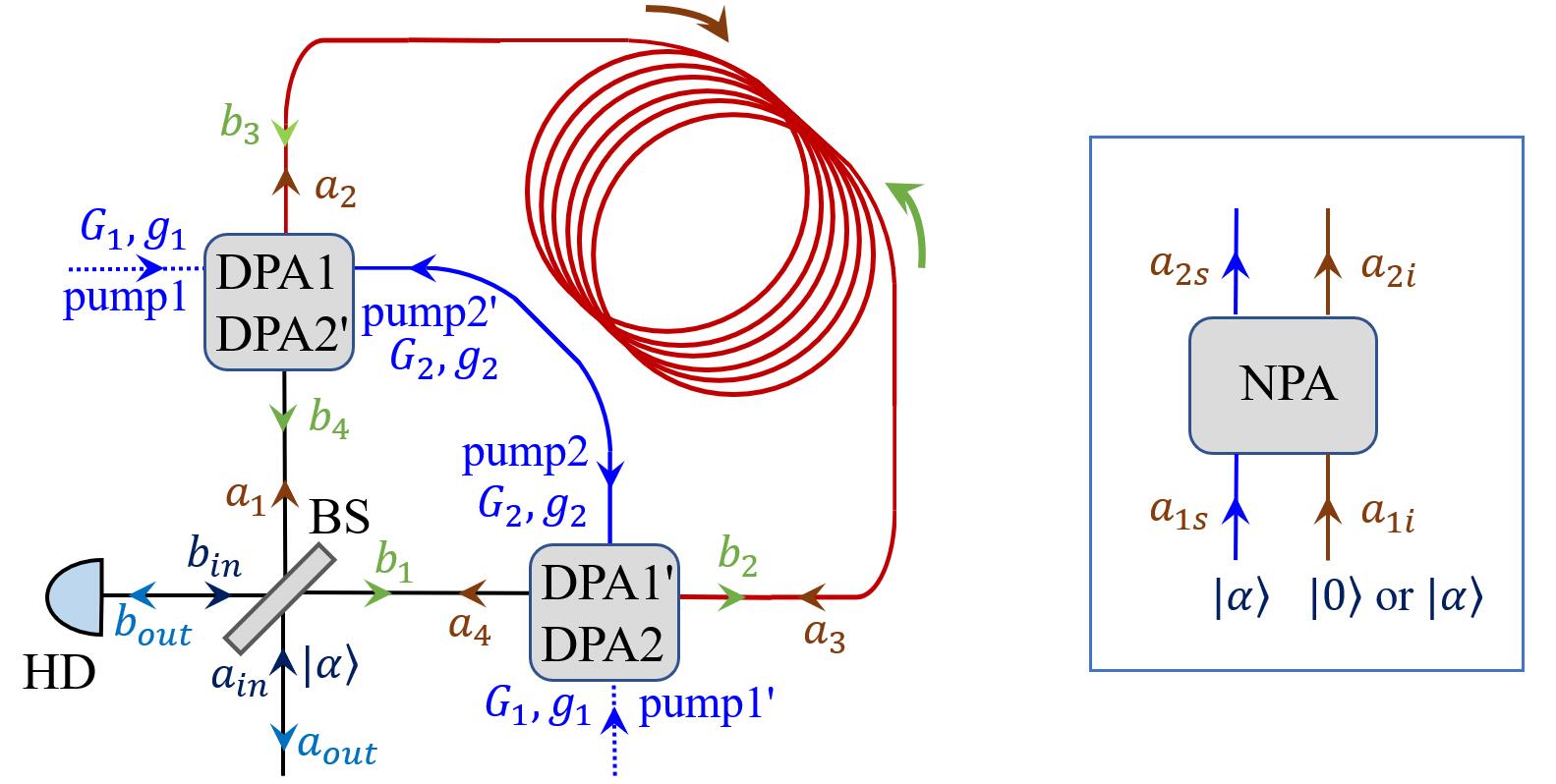}
\caption{A modified Sagnac interferometer with SU(1,1) interferometers nested inside. Inset: non-degenerate parametric amplifier (NPA) to replace degenerate parametric amplifiers (DPA). $s$: signal field, $i$: idler field.}
\label{setup2}
\end{figure}

Since SUI depends on phase sum, we need to send both signal and idler entangled fields generated in parametric amplifiers in the same direction for sensing the phase shift, as demonstrated in the dual-beam sensing scheme of SU(1,1) interferometers \cite{dual}. This means that we need two SUIs, one in each direction, to form the Sagnac geometry. As shown in Fig.\ref{setup2}, we insert parametric amplifiers  in a Sagnac interferomter to form SU(1,1) interferometers (red paths) in both directions. For simplicity, we first consider degenerate parametric amplifiers (DPA1, DPA2). 

Referring to the notations in Fig.\ref{setup2} for this scheme, a coherent state $|\alpha\rangle$  at input port $a_{in}$ is split by a 50:50 beam splitter (BS) into two fields $a_1$ and $b_1$ which are amplified by degenerate parametric amplifiers DPA1, DPA1' pumped by pump 1 and pump 1', respectively. The amplified fields $a_2$ and $b_2$ propagate through a rotation sensing loop (red) in opposite directions ($a_3, b_3$) before being amplified again by DPA2 and DPA2' pumped by pump 2 and 2', respectively, which are pass-through fields of pump 1 and 1' via the pump path (blue) but in opposite directions. The amplification processes of DPA1 and DPA2 and of DPA1' and DPA2' form two SU(1,1) interferometers in opposite directions for rotation sensing. The outputs $a_4$, $b_4$ of the SUIs are recombined by the beam splitter to complete the Sagnac interferometer. We detect field $b_{out}$ by homodyne detection (HD) for rotational signal.   

For the two input fields $(a_{in}$ and $b_{in})$ entering the two ports of the interferometer.
The output fields $(a_1$ and $b_1)$ from the BS can be written as
\begin{eqnarray}\label{BS1}
a_{1}=(a_{in}+b_{in})/\sqrt{2},~~
b_{1}=(b_{in}-a_{in})/\sqrt{2}.
\end{eqnarray}
The two fields $a_1$ and $b_1$, after phase shifted by $\delta_{a1}$ and $\delta_{b1}$, respectively, then each pass through a degenerate parametric amplifier  DPA1 and DPA1', respectively. According to the input-output relations of the DPA, the output fields ($a_2$ and $b_2$) are given by
\begin{eqnarray}\label{DPA1}
a_{2}&=&G_1a_1 e^{i\delta_{a1}} +e^{i\varphi_{1}}g_1a_1^{\dagger}e^{-i\delta_{a1}}, \cr
b_{2}&=&G_1b_1 e^{i\delta_{b1}}+e^{i\varphi_{1}^{\prime}}g_1b_1^{\dagger}e^{-i\delta_{b1}},
\end{eqnarray}
where $\varphi_{1}$ and $\varphi_{1}'$ are related to the phases of the pump fields (pump1 and pump1'). $G_1, g_1, (G_1^2-g_1^2=1)$ are the gain parameters of DPA1 and DPA1' when passing through the DPAs for the first time.

Subsequently, the two fields $(a_2$ and $b_2)$ travel oppositely along a common path for rotation sensing, there will be a phase difference between these two counter-propagating fields due to Sagnac effect. The fields $(a_3$ and $b_3)$ carrying the phase change are expressed as
\begin{eqnarray}
a_{3}&=&a_2e^{i\delta_{1}}=(G_1a_1 e^{i\delta_{a1}} +e^{i\varphi_{1}}g_1a_1^{\dagger}e^{-i\delta_{a1}})e^{i\delta_{1}}, \nonumber\\
b_{3}&=&b_2e^{i\delta_{2}}=(G_1b_1 e^{i\delta_{b1}}+e^{i\varphi_{1}^{\prime}}g_1b_1^{\dagger}e^{-i\delta_{b1}})e^{i\delta_{2}},
\end{eqnarray}
where $\delta=\delta_1-\delta_2 \ll 1$ is the phase difference due to the Sagnac effect. The phase-shifted $a_3$ and $b_3$ are sent to DPA2 and DPA2' to form nested SU(1,1) interferomters (SUI). The output fields $a_4$ and $b_4$ of the SUIs are
\begin{eqnarray}
a_{4}&=&G_2a_3+e^{i\varphi_{2}}g_2a_3^{\dagger} =G^{a}_Ta_1e^{i\delta_{a1}}+g^a_{T}a_1^{\dagger}e^{-i\delta_{a1}},\nonumber\\
b_{4}&=&G_2b_3+e^{i\varphi_{2}^{\prime}}g_2b_3^{\dagger} =G^{b}_Tb_1e^{i\delta_{b1}}+g^{b}_Tb_1^{\dagger}e^{-i\delta_{b1}},
\label{2th-DPA-out}
\end{eqnarray}
with
\begin{eqnarray}
G_T^{a}&=&e^{i\delta_{1}}G_1G_2+e^{i(\varphi_{2}-\varphi_{1}-\delta_{1})}g_1g_2,\cr 
G_T^{b}&=&e^{i\delta_{2}}G_1G_2+e^{i(\varphi_{2}^{\prime}-\varphi_{1}^{\prime}-\delta_{2})}g_1g_2,\cr
 g_T^{a}&=&e^{i(\delta_{1}+\varphi_{1})}G_2g_1+e^{i(\varphi_{2}-\delta_{1})}G_1g_2, \cr
g_T^{b}&=&e^{i(\delta_{2}+\varphi_{1}^{\prime})}G_2g_1+e^{i(\varphi_{2}^{\prime}-\delta_{2})}G_1g_2,
\end{eqnarray}
where $\varphi_{2}$ and $\varphi_{2}'$ are the phases from pump2 and pump2', $G_2, g_2, (G_2^2-g_2^2=1)$ are the gain parameters of DPA2 and DPA2'. Since pump 2 (2') is directly from pump 1 (1'), we have  $G_1=G_2\equiv G, g_1=g_2\equiv g$ with $G^2-g^2=1$. The interferometer works best around the dark fringe with $\varphi_{1}=\varphi_{1}^{\prime}=0$, $\varphi_{2}=\pi+\delta_{\varphi_2}$ and $\varphi_{2}^{\prime}=\pi+\delta_{\varphi_2^{\prime}}$ where $\delta_{\varphi_2}$ and $\delta_{\varphi_2'}$ are small phase shifts on the pump fields caused by the rotation. Under these conditions,
\begin{eqnarray}\label{apprx}
G_T^{a}&\approx& 1+i\delta_{1}(G^2+g^2)-i\delta_{\varphi_2}g^2, ~~
g_T^{a}\approx iGg (2\delta_{1}- \delta_{\varphi_2}), \cr
G_T^{b}&\approx& 1 +i\delta_{2}(G^2+g^2)-i\delta_{\varphi_2'}g^2, \cr
g_T^{b}&\approx & i Gg(2\delta_{2}-\delta_{\varphi_2'}).
\end{eqnarray}

Combining the two output fields from DPAs, the output fields from the Sagnac-SU(1,1) interferometer are
\begin{eqnarray}
a_{out}&=&(a_4e^{i\delta_{a4}}-b_4e^{i\delta_{b4}})/\sqrt{2}, \cr
b_{out}&=&(b_4e^{i\delta_{b4}}+a_4e^{i\delta_{a4}})/\sqrt{2}.
\label{SUI-out}
\end{eqnarray}
From Eqs.(\ref{2th-DPA-out} - \ref{SUI-out}), we obtain
\begin{eqnarray}
a_{out}&=&[(G^{a}_Ta_1e^{i\delta_{a1}}+g^a_{T}a_1^{\dagger}e^{-i\delta_{a1}})e^{i\delta_{a4}}\cr &&\hskip 0.4in -(G^{b}_Tb_1e^{i\delta_{b1}}+g^{b}_Tb_1^{\dagger}e^{-i\delta_{b1}})e^{i\delta_{b4}}]/\sqrt{2}  \cr
&=&(\lambda_1 b_{in}+ \lambda_2 a_{in}+\lambda_3 b_{in}^{\dagger}+\lambda_4 a_{in}^{\dagger})/2,\cr
b_{out}&=&[(G^{b}_Tb_1e^{i\delta_{b1}}+g^{b}_Tb_1^{\dagger}e^{-i\delta_{b1}})e^{i\delta_{b4}}\cr &&\hskip 0.4in+(G^{a}_Ta_1e^{i\delta_{a1}}+g^a_{T}a_1^{\dagger}e^{-i\delta_{a1}})e^{i\delta_{a4}}]/\sqrt{2} \cr
&=&(\lambda_1 a_{in}+\lambda_2 b_{in}+\lambda_3 a_{in}^{\dagger}+\lambda_4 b_{in}^{\dagger})/2,
\end{eqnarray}
where
\begin{eqnarray}
\lambda_1&=&G^{a}_Te^{i(\delta_{a1}+\delta_{a4})}-G^{b}_Te^{i(\delta_{b1}+\delta_{b4})},\cr
\lambda_2&=&G^{a}_Te^{i(\delta_{a1}+\delta_{a4})}+G^{b}_Te^{i(\delta_{b1}+\delta_{b4})},\cr
\lambda_3&=&g^a_{T}e^{i(\delta_{a4}-\delta_{a1})}-g^{b}_Te^{i(\delta_{b4}-\delta_{b1})},\cr
\lambda_4&=&g^a_{T}e^{i(\delta_{a4}-\delta_{a1})}+g^{b}_Te^{i(\delta_{b4}-\delta_{b1})}.
\end{eqnarray}

For the counter-propagation fields, Sagnec effect gives $\delta_1=-\delta_2\equiv\delta/2$, $\delta_{a1}=-\delta_{b4}\equiv\Delta_1/2$, $\delta_{a4}=-\delta_{b1}\equiv\Delta_2/2$
 and $\delta_{\varphi_2}=-\delta_{\varphi_2'}\equiv\Delta_{\varphi_2}/2$. Using the operation condition and the approximations leading to Eq.(\ref{apprx}), we obtain
\begin{eqnarray}
\lambda_1&\approx & i[\Delta_1+\Delta_2+ \delta(G^2+g^2) -\Delta_{\varphi_2}g^2], \nonumber\\
\lambda_2&\approx & 2, ~~~
\lambda_3 \approx   iGg(2\delta -i\Delta_{\varphi_2}), ~~~
\lambda_4 \approx  0,
\end{eqnarray}
where we dropped the higher order terms. The output fields of the modified Sagnac interferometer are then
\begin{eqnarray}\label{dg-out}
a_{out}&\approx& a_{in} + iGg(2\delta -\Delta_{\varphi_2})b_{in}^{\dagger}/2 \cr 
&& \hskip 0.2in   + i[\Delta_1+\Delta_2+ \delta(G^2+g^2) -\Delta_{\varphi_2}g^2] b_{in}/2\nonumber\\
b_{out}&\approx& b_{in} + iGg(2\delta -\Delta_{\varphi_2})a_{in}^{\dagger}/2 \cr 
&& \hskip 0.2in   + i[\Delta_1+\Delta_2+ \delta(G^2+g^2) -\Delta_{\varphi_2}g^2]a_{in}/2.~~~~~~
\end{eqnarray}

\begin{figure}
\centering
\includegraphics[width=0.95\linewidth]{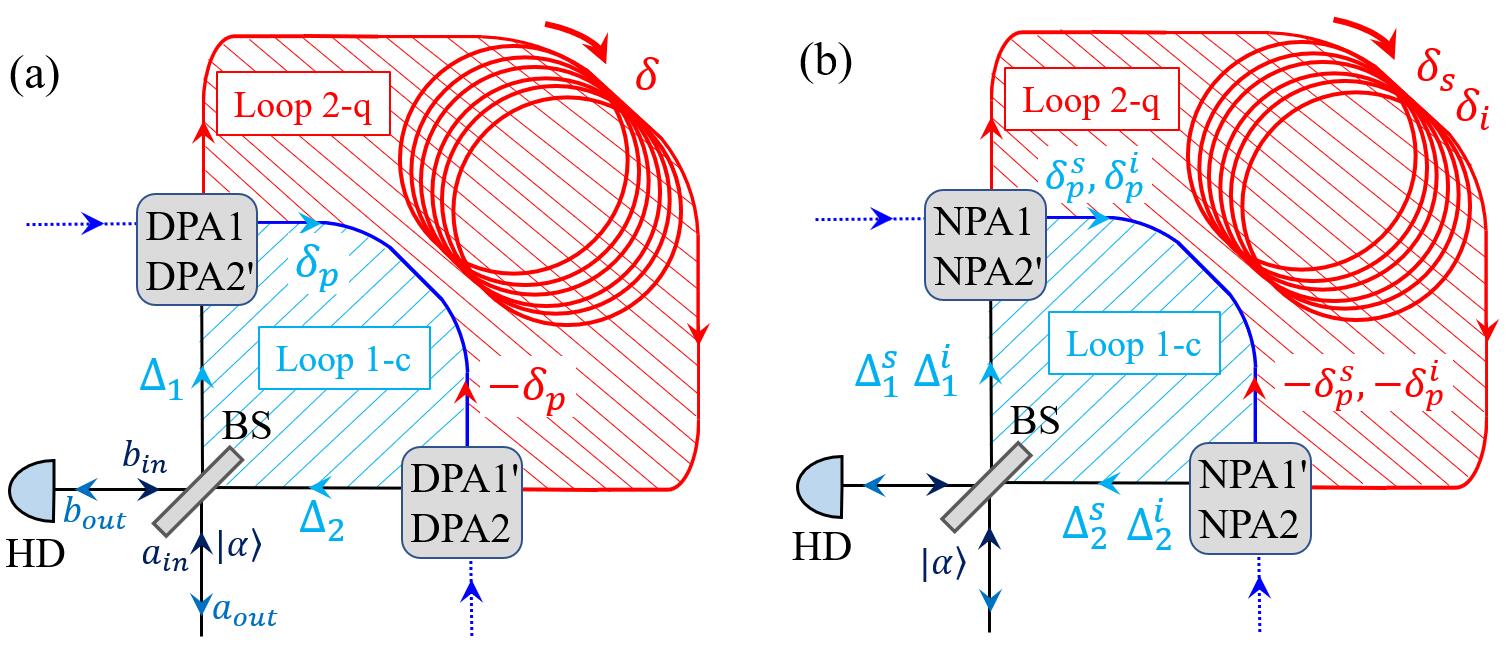}
\caption{Modified Sagnac interferometer with SU(1,1) interferometers nested inside. (a) Degenerate case (b) Non-degenerate case. Light blue shaded area is Loop 1-c, which is for classical fields and red shaded area is Loop 2-q for quantum entangled fields. Arrows refer to loop directions.}
\label{loops}
\end{figure}

For a coherent state $|\alpha\rangle$ input at $a_{in}$ with positive $\alpha =|\alpha|$ for simplicity and vacuum at $b_{in}$, when there is no rotation, the output of $b_{out}$ is dark in vacuum and all input energy is out from $a_{out}$. With rotation, the output of $b_{out}$ contains only the rotation-induced phase shift information. So, we measure the quadrature-phase amplitude of the dark port output field $b_{out}$ by homodyne detection (HD), which takes the form of $Y_{b_{out}} = i(b_{out}^{\dagger}-b_{out})$ for phase.
The noise fluctuation of $Y_{b_{out}}$ is mostly from the first term $b_{in}$ in Eq.(\ref{dg-out}), which is in  vacuum:
\begin{eqnarray}
\langle \Delta Y_{b_{out}}^2\rangle_N
\approx 1,
\label{noise}
\end{eqnarray}
whereas the signal power of $Y_{b_{out}}$ is
\begin{eqnarray}
\langle Y_{b_{out}}\rangle^2_S &=& |\alpha|^2[\Delta_1+\Delta_2+(G+g)^2\delta \cr
&& \hskip 0.7 in -\Delta_{\varphi_2} g(G+g)]^2.
\label{signal}
\end{eqnarray}
Then we can find the signal-to-noise ratio (SNR):
\begin{eqnarray}\label{SNR}
	SNR_{Y_{bout}}&\equiv &\frac{\langle Y_{b_{out}}\rangle^2_S }{\langle \Delta Y_{b_{out}}^2\rangle_N} =  |\alpha|^2[\Delta_1+\Delta_2+(G+g)^2\delta \cr
&& \hskip 1 in -\Delta_{\varphi_2} g(G+g)]^2.
\end{eqnarray}
Defining $\delta_p\equiv \Delta_{\varphi_2}/2$ as the equivalent phase shift at degenerate frequency of signal and idler induced by rotation in the pump path (dark blue curve in Fig.\ref{loops}a), we have
\begin{eqnarray}\label{SNR1}
SNR_{Y_{bout}}&=&
|\alpha|^2[\Delta_1+\Delta_2+(G+g)^2\delta -2\delta_{p} g(G+g)]^2\cr
&=&|\alpha|^2[\Delta_1+\Delta_2+(G+g)^2(\delta-\delta_p)\cr
&& \hskip 1in +(G+g)(G-g)\delta_{p}]^2\cr
&=&|\alpha|^2[(\Delta_1+\Delta_2+\delta_{p})+(G+g)^2(\delta-\delta_p)]^2\cr
&=&[\Delta_{lp1-c}+\Delta_{lp2-q}(G+g)^2]^2|\alpha|^2,
\end{eqnarray}                                                                     
where we use $G^2-g^2=1$ and as shown in Fig.\ref{loops}(a), $\Delta_{lp1-c}\equiv \Delta_1+\delta_{p}+\Delta_2$ is the phase shift associated with the first loop (Loop 1-c of light blue area $A_{lp1-c}$) including only classical fields of the pump fields and the fields before parametric amplifiers so that $\Delta_{lp1-c} = 8\pi \Omega A_{lp1-c}/\lambda c$ for angular velocity $\Omega$ of rotation ($\lambda =$ the wavelength of the rotation sensing fields) \cite{sag},  whereas aftering noting that $-\delta_p$ is in opposite direction of $\delta_p$ in the loops, $\Delta_{lp2-q}\equiv \delta + (-\delta_p)$ is the phase shift associated with the second loop (Loop 2-q of red area $A_{lp2-q}$) involving quantum entangled fields inside the SU(1,1) interferometer so that $\Delta_{lp2-q} = 8\pi \Omega A_{lp2-q}/c \lambda$. Making the classical loop area (Loop 1-c) negligible, we have
\begin{eqnarray}\label{SNR2}
	SNR_{Y_{bout}}&\approx &\Delta_{lp2-q}^2(G+g)^2I_{ps} \cr &=& 64\pi^2 \Omega^2 A^2_{lp2-q} (G+g)^2I_{ps}/\lambda^2 c^2,
\end{eqnarray}    
where $I_{ps} \equiv (G+g)^2 |\alpha|^2$  is the total phase sensing photon number. Notice that when there is no parametric amplifier so that  $G=1$ and $g=0$, the SNRs in both Eqs. (\ref{SNR1},\ref{SNR2}) becomes that at shot noise limit for a regular Sagnac interferometer \cite{lin79}. This result shows that there is an enhancement factor of $(G+g)^2$  in SNR. 

Degenerate parametric amplifiers (DPA) are phase sensitive and require phase lock of the pump fields. This is inconvenient experimentally. We can replace DPAs with phase insensitive non-degenerate parametric amplifiers (NPA), as shown in the inset of Fig.\ref{setup2}. The difference is that we now deal with two fields of signal and idler, which may have different frequencies, instead of just one field and the amplifier relations in Eqs.(\ref{DPA1}, \ref{2th-DPA-out}) change to 
\begin{eqnarray}\label{NPA1}
a_{2s}&=&G_1a_{1s} e^{i\delta_{a1s}} +e^{i\varphi_{1}}g_1a_{1i}^{\dagger}e^{-i\delta_{a1i}}, \cr
a_{2i}&=&G_1a_{1i} e^{i\delta_{a1i}}+e^{i\varphi_{1}}g_1a_{1s}^{\dagger}e^{-i\delta_{a1s}},
\end{eqnarray}
for non-degenerate parametric amplifiers NPA1, NPA1', and 
\begin{eqnarray}\label{NPA2}
a_{4s}=G_2a_{3s} +e^{i\varphi_{2}}g_2a_{3i}^{\dagger}, ~a_{4i}=G_2a_{3i} +e^{i\varphi_{2}}g_2a_{3s}^{\dagger}~~~~~~
\end{eqnarray}
for NPA2 and NPA2'.  
For the $b$-fields in the opposite direction, we just need to change $a$ to $b$ and $\varphi_{1,2}$ to $\varphi_{1,2}^{\prime}$ in the expressions above. The BS relations in Eq.(\ref{BS1}, \ref{SUI-out}) also need to add labels ``$s$" and ``$i$" for signal and idler fields, respectively. Using the operation condition and small rotation induced shift approximation as before, we obtain the output fields of the modified Sagnac interferometer but here only present $b_{out}$-field for detection:
\begin{eqnarray}
b_{out}^{(s)}&\approx& b_{in}^{(s)}  +iGg(\delta_s +\delta_i  - \Delta_{\varphi_2}) a_{in}^{(i)\dagger}/2\cr 
&&\hskip 0.1in + i [\Delta_{1}^s+\Delta_{2}^s+\delta_s G^2+g^2 (\delta_i  - \Delta_{\varphi_2}) ]a_{in}^{(s)}/2 \cr 
b_{out}^{(i)}&\approx& b_{in}^{(i)}  +iGg(\delta_i +\delta_s - \Delta_{\varphi_2}) a_{in}^{(s)\dagger}/2\cr 
&&\hskip 0.1in + i [\Delta_{1}^i+\Delta_{2}^i+\delta_i G^2+g^2 (\delta_s  - \Delta_{\varphi_2}) ]a_{in}^{(i)}/2,~~~~~~
\end{eqnarray}
where $\delta_s\equiv \delta_{s1} - \delta_{s2}, \delta_i\equiv \delta_{i1} - \delta_{i2}$ are the rotation induced phase shifts for signal and idler fields, respectively and $\Delta_{1,2}^{s,i}$ are the phases of the injecting signal and idler fields before NPAs (Fig.\ref{loops}(b)).

With a coherent state $|\alpha\rangle$ input at $a^{(s)}_{in}$ but vacuum $|0\rangle$ at $a^{(i)}_{in}$, the gains of the NPAs are phase insensitive. We make measurement of $Y_{b^{(s)}_{out}} = (b^{(s)}_{out}-{b}_{out}^{{(s)}\dag})/i$ for phase information. By going through the similar derivation leading to Eq.(\ref{SNR}), we have 
\begin{eqnarray}\label{SNR-NPA1}
	{SNR}_{b^{(s)}_{out}}
=|\alpha|^2[\Delta_{1}^s+\Delta_{2}^s+\delta_{s}G^2+\delta_{i}g^2-\Delta_{\varphi_2} g^2]^2.~~~~~~
\end{eqnarray}  
Since the pump phase is related to the phases of signal and idler fields due to energy conservation ($\omega_p~{\rm or}~ 2\omega_p =\omega_s+\omega_i$) by $\Delta_{\varphi_2} = \delta^s_p+\delta_p^i$ with $\delta^{s,i}_p$ as the equivalent signal or idler phase for the pump path (dark blue curve in Fig.\ref{loops}b), we can write further similar to Eq.(\ref{SNR1}):
\begin{eqnarray}\label{SNR-NPA2}
	SNR_{b^{(s)}_{out}}
=[\Delta_{lp1-c}^{(s)}+\Delta_{lp2-q}^{(s)}G^2+\Delta_{lp2-q}^{(i)}g^2]^2|\alpha|^2,~~~~~~
\end{eqnarray}  
where $\Delta_{lp1-c}^{(s)}\equiv \Delta_{1}^s+\Delta_{2}^s+\delta^s_p=8\pi \Omega A_{lp1-c}/\lambda_s c$ is for the classical loop (light blue) and $\Delta_{lp2-q}^{(s,i)}$ $\equiv \delta_{s,i} +(- \delta^{s,i}_p) = 8\pi \Omega A_{lp2-q}/\lambda_{s,i} c$ for the quantum loop (red).  Note that the enhancement factors are different for signal and idler phase shifts. This is due to their different strengths. 

For the frequency near degenerate case of $\lambda_{s}\approx \lambda_{i}\equiv \lambda$, we have $\Delta_{lp2-q}^{(s)} \approx \Delta_{lp2-q}^{(i)}\equiv \Delta_{lp2-q}$ and if we make the classical loop as small as possible, Eq.(\ref{SNR-NPA2}) becomes
\begin{eqnarray}\label{SNR-NPA3}
	SNR_{b^{(s)}_{out}} =\Delta_{lp2-q}^2(G^2+g^2)I_{ps}^{(N)},
\end{eqnarray}  
where $I_{ps}^{(N)} \equiv (G^2+g^2)|\alpha|^2$ is the total phase sensing photon number in the non-degenerate case. Note the quantum enhancement factor is $G^2+g^2$, which is about half of that for degenerate case when $G\approx g$. This is because we have vacuum injection to the idler modes (see inset of Fig.\ref{setup2}). To avoid this, we can also input a coherent state at the idler field with equal strength as the signal field, that is, $|\alpha\rangle$ for both $a^{(s)}_{in}$ and $a^{(i)}_{in}$. Following the same procedures as before, it is straightforward to find
\begin{eqnarray}\label{SNR-NPAsi}
	SNR_{b^{(s,i)}_{out}}
&=&[\Delta_{1}^{s,i}+\Delta_{2}^{s,i}+(G+g)(\delta_{s,i}G+\delta_{i,s}g)\cr 
&& \hskip 0.2in -(\delta^{s}_p+\delta^{i}_p) g(G+g)]^2 |\alpha|^2\cr
&=& [\Delta_{lp1-c}^{(s,i)}+\Delta_{lp2-q}^{(s,i)}(G+g)^2\cr
&&\hskip 0.2in +(\Delta_{lp2-q}^{(i,s)}-\Delta_{lp2-q}^{(s,i)})g(G+g)]^2|\alpha|^2,~~~~
\end{eqnarray}  
which is similar to the degenerate case of Eq.(\ref{SNR1}) for near degenerate case of $\lambda_s\approx \lambda_i$ when the last term is approximately zero. But now the total input photon number is $2|\alpha|^2$ due to input from both signal and idler fields so the SNR is halfed. On the other hand, both signal and idler output fields contain the phase shift information so we need to combine them by measuring $Y_{b^{(s)}_{out}} + Y_{b^{(i)}_{out}}$. For this, the noise power doubles but the signal sizes from the two add before taking square, leading to 
\begin{eqnarray}\label{SNR-NPAsi2}
	SNR_{b^{(s)}_{out}+b^{(i)}_{out}}
&=& [\Delta_{lp1-c}^{(s)}+\Delta_{lp1-c}^{(i)}\cr
&&\hskip 0.1in +(\Delta_{lp2-q}^{(s)}+\Delta_{lp2-q}^{(i)})(G+g)^2]^2|\alpha|^2/2\cr
&=& (2|\alpha|^2) [\bar{\Delta}_{lp1-c}+\bar{\Delta}_{lp2-q}(G+g)^2]^2,~~~~~~
\end{eqnarray}  
where $\bar{\Delta}_{lp1-c} = (\Delta_{lp1-c}^{(s)}+\Delta_{lp1-c}^{(i)})/2, \bar{\Delta}_{lp2-q}=(\bar{\Delta}_{lp2-q}^{(s)}+\bar{\Delta}_{lp2-q}^{(i)})/2$ are the average quantities of the signal and idler fields. Using energy conservation $\omega_s+\omega_i = \omega_p$ $(\rm or~ 2\omega_p)$ of parametric process in $\chi^{(2)}$ (or  $\chi^{(3)}$ ) medium, we have $\bar{\Delta}_{lp1-c} = 8\pi \Omega A_{lp1-c}/c \lambda_p (\rm or~\lambda_p/2)$ and $\Delta_{lp2-q} = 8\pi \Omega A_{lp2-q}/c \lambda_p (\rm or~\lambda_p/2)$, which only depend on the pump wave length $\lambda_p$. This result is the same as Eq.(\ref{SNR1}) for the degenerate case with a total input photon number of $2|\alpha|^2$. Therefore, we obtain the same enhancement factor of $(G+g)^2$ as the degenerate case. 

Likewise in the single injection case of Eq.(\ref{SNR-NPA2}), we can also make use of the idler output which contains the phase information due to rotation as well. It is straightforward to show that at large gain of $G\approx g \gg 1$, the SNR will have the same enhancement factor of $(G+g)^2$ as the degenerate case.

Notice that the equivalent pump phase difference $-\delta_p$ (or $-\delta^{s}_p,-\delta^{i}_p$ in the non-degenerate case) is also involved in the quantum part $\Delta_{lp2-q} = \delta + (-\delta_p)$ of the phase shift in addition to $\delta$ of the entangled quantum fields. This is so because the total phase of the SU(1,1) interferomters involves the phase of the pump fields. Therefore, $\Delta_{lp2-q}$ is the overall phase shift from SU(1,1) interferometers. In practice, the pump fields are usually strong and may cause some detrimental thermal effects so we need to make $\delta_p$ as small as possible so that $\Delta_{lp2-q} \approx \delta$ and only the phase shifts from quantum fields are involved.

In conclusion, we showed that we can modify the Sagnac interferometer by inserting SU(1,1) interferometers inside. This can achieve quantum enhanced sensitivity in rotational sensing or reduce the sensing power to avoid thermal effects but reach the same sensitivity as the equivalent classical Sagnac interferometer. 

\section*{AUTHOR DECLARATIONS}
The authors have no conflicts to disclose.

\begin{acknowledgments}
This work was supported in part by. National Natural Science Foundation of China (Grants No. 91836302, No. 12004279, and No. 12074283) and by General Research Fund from Hong Kong Research Grants Council (No. 11315822)
\end{acknowledgments}

\section*{Data Availability Statement}
Data sharing is not applicable to this article as no new data were created or analyzed in this study.

\

\end{document}